\documentstyle[11pt,aaspp4]{article}

\begin{document}

\title{Radial Velocity Studies of Close Binary Stars.~I}

\author{Wenxian Lu and Slavek M. Rucinski\\
e-mail: {\it lu@astro.utoronto.ca, rucinski@astro.utoronto.ca\/}}
\affil{David Dunlap Observatory, University of Toronto \\
P.O.Box 360, Richmond Hill, Ontario L4C~4Y6, Canada}

\centerline{\today}

\begin{abstract}
Radial velocity data are presented for 10 W~UMa-type systems:
GZ~And, V417~Aql, LS~Del, EF~Dra, V829~Her, FG~Hya, AP~Leo,
UV~Lyn, BB~Peg, AQ~Psc, together with preliminary,
circular-orbit determinations of spectroscopic elements,
with the main goal of obtaining mean radial
velocities and mass-ratios for these systems.
This is the first part of a series which
will contain radial velocity data for Northern-hemisphere, 
short-period eclipsing binaries, accessible to medium-resolution 
spectroscopic studies with 1.8-meter class telescopes.
\end{abstract} 

\keywords{ stars: close binaries - stars: eclipsing binaries -- 
stars: variable stars}

\section{INTRODUCTION}

Contact binary stars consisting of solar type components (also called 
W~UMa-type binary stars) are very common: According to the new,
unbiased statistics, a by-product of the OGLE microlensing project, their 
{\it apparent\/} 
frequency of occurrence among main sequence stars of spectral types 
F to K (intrinsic colors $0.4 < V-I_C < 1.4$) is about $1/130 - 1/100$,
which leads to the spatial frequency at the level
$1/60 - 1/80$ (Rucinski \markcite{ruc98}
1998). Most of them have orbital periods within
$0.25 < P < 0.7$ days. They apparently do not exist below the
orbital period of 0.22 days and they are rare at orbital
periods longer than 0.7 days and spectral types earlier than
about F0 to F2. The kinematic data for contact 
systems in the solar neighborhood (Guinan \& Bradstreet \markcite{gb88}
1988) and the spatial distribution in the Baade's Window direction
(Rucinski \markcite{ruc97a} \markcite{ruc97b} 1997a, 1997b = R97a, R97b),
suggest an Old Disk population of interacting binaries formed primarily during
the Turn-Off-Point stage of evolution. This stage is conductive 
to rapid synchronization and formation of contact systems 
from close, but detached binaries when component stars expand and
are able to interact to the point of mutual contact.

While survey data such as from the OGLE project can give statistical
information on the relative numbers of stars in various stages of 
evolution and on duration of the pre- and in-contact stages, only
detailed studies of individual systems in the solar neighborhood
can yield absolute parameters of the systems and can help in
establishing correct kinematic properties of this contact binary
population. Also, precise data for individual bright contact systems
should lead to improvements in the luminosity calibration and in
their use as distance indicators (R97a, R97b),
These aspects formed the main rationale of a program started
by the current paper of the series.

The Hipparcos astrometric satellite provided the first set of
reliable parallaxes and proper motions for a sample of contact
binaries: Parallaxes giving error in absolute magnitude $\epsilon M_V
< 0.5$ became available for 40 systems (Rucinski \& Duerbeck 
\markcite{rd97} 1997),
with half that number having errors $\epsilon M_V < 0.25$. Most of these 
systems have high-quality proper-motion determinations with errors
at the level of $1 - 10$ milli-arcsec per year. 
The proper motions data can be combined with
mean systemic radial velocities $\gamma$ to provide spatial velocities of these
stars. However, many systems with excellent parallax and tangential-velocity
data from Hipparcos lack correspondingly accurate radial velocity data.
The present paper describes observations of 10 systems 
collected over the recent years
at the Dominion Astrophysical Observatory and at the David Dunlap 
Observatory (DDO), together with results of a new program which was 
initiated at the DDO to determine primarily the mean radial 
velocities $\gamma$. It is hoped that the next papers of this series will 
be more homgeneous in that they will contain the DDO results only;
they will describe typically ten systems per paper. 
 
Determinations of $\gamma$ velocities cannot be
usually separated from determinations of the velocity amplitudes, $K_1$ and
$K_2$. Thus, the important product of this program are also the
spectroscopic mass-ratios, $q = K_2/K_1$, as well as individual masses
of the components. For determinations of stellar masses, orbital inclination
angles are needed, but these can only come from light curve solutions. In
turn, light curve solutions gain in quality when spectroscopic values
of $q$ are used. Although we did recognize that combined spectroscopic
-- photometric solutions would provide the most
consistent approach, we decided not to perform full solutions of our
program binaries through combination of the new velocity data with the
light curves from the literature. The main reason for this decision was
that light curves were simply not available for many of our systems or that
in some cases they did exist, but were of disputable 
quality. Thus, the goal of these observations was
simply to provide radial velocity data and 
the simplest possible circular solutions giving $\gamma$, $q$, $K_1$ and
$K_2$. The selection of the objects was based on their brightness, 
visibility from the northern hemisphere and availability 
of astrometric data from the Hipparcos satellite. It is hoped that this
program will lead eventually spectroscopic data for all short period
and contact binaries accessible to 2-meter class Northern-hemisphere
telescopes equipped with spectrographs providing spectral resolutions of 
the order of 10,000 -- 15,000.

\section{OBSERVATIONS AND DATA REDUCTIONS}

Observations were collected at the Dominion Astrophysical
Observatory (DAO), Victoria, B.C.\ for 3 stars (LS~Del, AP~Leo, UV~Lyn)
and at the David Dunlap Observatory (DDO) near 
Toronto for the remaining 7 stars. 
At DAO, the 1.8-meter telescope and the
``21121'' spectrograph giving dispersion of 15 \AA/mm were used. 
The CCD detectors, central wavelengths 
and spectral windows were not the same so 
that they are described below in discussions of individual systems. Typically,
about 240 \AA\ were covered at 0.23 to 0.25 \AA/pixel at 4160 or 5020 \AA.
At DDO, the 1.9-meter telescope and the Cassegrain spectrograph illuminating
a CCD at a dispersion of 10.8 \AA/mm corresponding to about 
0.2 \AA/pixel or about 12 km/s/pixel was used for almost
all observations (see also the descriptions for individual systems below).
All the spectra for stars observed at DDO were centered at 
5185 \AA\ giving the spectrum coverage of 210 \AA.

The velocity determinations of binary component stars were done using
cross-correlation of spectra with standard stars (program VCROSS,
Hill \markcite{hil82} 1982) or broadening-function (BF) algorithms 
(Rucinski \markcite{ruc92} 1992), which also use 
spectra of sharp-line standard stars. The former
method was used for the systems observed at DAO, while the latter method
was used for the systems observed at DDO. In both cases,
the radial velocities of stars were determined by fitting Gaussians to the 
peaks. Tests showed that systematic differences between VCROSS and BF results
were well below the level of formal errors so that it 
was decided to disregard this minor inconsistency in the methodology. 

Spectra of different stars were used 
as sharp-line templates for cross-correlation and 
broadening-function determinations. They were, for 
systems observed at the DAO:
HD~32963 (G2V) for LS~Del, HD~126053 (G1V) for AP~Leo, 
UV~Lyn (HD~22484 (F8V) and HD~32963 (G2V));
at the DDO, HD~22484 (F8V) was used for AQ~Psc, V417~Aql, BB~Peg,
GZ~And, and FG~Hya, whereas  HD~102870 (F9V) was used for EF~Dra and
HD~187691 (F8V) was used for V829~Her.

The spectra of the program stars were reduced with 
IRAF\footnote{IRAF is distributed by the National Optical
Astronomy Observatories, which are operated by the Association of
Universities for Research in Astronomy, Inc., under cooperative
agreement with the National Science Foundation.}. The standard
procedures were employed, which consisting of de-biasing, cosmic-ray
removal, flat-fielding, extraction of one-dimensional spectra,
wavelength calibration, and rectification. 
The radial velocity orbits were solved using programs RVORBIT by
Graham Hill (1986, private communication) 
and SBCM by Chris Morbey \markcite{mor75} (1975). 
For each system, circular solutions of the form:
$V(t) = \gamma + K_i \sin (\phi)$, with $\phi$ being the phase,
were made separately for each component and then
for the whole system. No significant differences in $\gamma$
velocities for separate solutions were noted in any of the ten
systems. Normally, unit weights were given to individual observations
with exceptions of those indicating
blending at eclipses. The initial epoch $T_0$ was
included among unknowns in all cases due to lack of recent photometric
eclipse timings or possibilities of period changes. The values of $T_0$
which we give correspond to moments of the deeper eclipses for
each of the systems.

Standard stars were observed on each observing night of program stars. These
observations were used to check zero points in velocity. For DAO and early
DDO observations, zero points were accurate to 1.2 km/sec on average, while
for the 1996-1997 DDO observations, the accuracy (limited by the spectrograph 
flexure) was slightly lower and equaled 1.6 km/sec. 

From a series of over 60 observations of IAU standard stars,
it was found that the velocity of the template-star HD~22484, 
which was used for the radial velocity determinations of the DDO observations, 
had the measured value by 1.8 km/sec more negative than the IAU value.
The effect may have been caused by a nightly instability of the 
Cassegrain spectrograph. Thus, the derived radial velocities for five
systems, AQ Psc, V417 Aql, BB Peg, GZ And, and FG Hya,  
should be increased by 1.8 km/sec, leading to a corresponding
increase in the derived $\gamma$ velocity for each of these systems.
This value of 1.8 km/s may be taken as an estimate of the 
external error of our observations.

The observations are listed in Table~\ref{tab1} listing the heliocentric
Julian Day of the exposure, the adopted phase and the two radial 
velocity determinations, together with the deviations from the circular
solution. The stars are arranged 
in this table in the variable-star name order, 
that is by the alphabetic order of constellations. 
The next section discusses individual systems. For each system, 
we first describe the observations and then
summarize briefly how they relate to the previous existing data. 
For some quantities we quote determination uncertainties; they are
given in parentheses, following the actual numbers. This convention is used
in Table~\ref{tab2} which summarizes the results of the
paper. Figures~\ref{fig1} -- \ref{fig3}
 show individual observations for the program
stars together with the respective sinusoidal fits.

\placetable{tab1}

\placetable{tab2}

\section{RESULTS FOR INDIVIDUAL SYSTEMS}

\subsection{GZ And}

The spectroscopic orbit presented here, and based on the DDO observations
has been determined for the first time.

A total of 27 spectra were obtained in the period of
September -- December 1996. The exposure times were 15 minutes
corresponding to about 0.034 of the orbital period. This time resolution
is a somewhat lower than for the other systems due
to the shortness of the period. The orbital period of
0.3050067 day has been adopted after Liu et al.\ \markcite{liu87} (1987).

The system was discovered as a W~UMa-type binary by  
Walker \markcite{wal73} (1973), as a 
brightest component (A) of a visual multiple
system ADS~1693. whose properties were studied later by Walker 
\markcite{wal91} \markcite{wal96} (1991, 1996). 
He claimed that, additionally,
the W~UMa binary belongs to a close triple system with the period of
about 5.3 years for the wide pair. Indeed, on the basis of the photometric 
properties it appears that all 
visual components of ADS~1693, except star D,
are probably physical members of the multiple system.
However, we have been unable to detect the 
third component in our spectroscopic 
observations of GZ~And (ADS~1963~A). 
Both Walker \markcite{wal91} \markcite{wal96} 
(1991, 1996) and Liu et al.\ \markcite{liu87} (1987) signaled that they made 
photometric solutions, but their results are not
available yet, so that this W-type system still awaits a combined
photometric -- radial velocity investigation.

\subsection{V417 Aql}

The spectroscopic orbit presented here has been determined for 
the first time. The system was observed at DDO
in August -- September 1996. A total of 22 spectra 
were obtained, all of them at phases around the orbital
quadratures. The exposure times were 15 minutes long,
corresponding to a phase resolution of 0.028. A 
period of 0.37031142 was adopted in the spectroscopic orbit solution,
following Faulkner \markcite{fau86} (1986).

V417 Aql was discovered by Hoffmeister \markcite{hof35} (1935). 
There were few investigations of this system other than those devoted to 
timing of eclipse minima. Recently, Samec et al.\ \markcite{sam97} (1997) 
obtained high-precision photometric light curves and
made solutions of geometric elements. Their mass ratio of 
$q=0.3684$ is almost the same as our spectroscopic value of 
0.362. By combining the photometric solution of Samec et al.\ with
our radial velocity solution, the following absolute parameters
can be derived: $a=2.68\,R_\odot$, side radii $R_1=1.29\,R_\odot$, 
and $R_2 = 0.80\,R_\odot$, and masses $M_1=1.40\, M_\odot$, 
$M_2=0.50\,M_\odot$. The degree of contact is 19\%.
This W-type system is very similar to BB~Peg which we discuss below.

\subsection{LS~Del}

The spectroscopic orbit for this system is presented here
for the first time. The star was observed at 
DAO with the ``21121'' spectrograph and the
RCA2 CCD detector at the dispersion of 15 \AA/mm, or the
scale 0.25 \AA/pixel. The binary was observed only on two nights, 
September 19 and 21, 1990. The spectra were centered at 4160 \AA\ 
with a coverage of 250 \AA. 42 observations were obtained, of 
which 29 were at orbital quadratures, and 13 were at
conjunctions. The conjunctions were observed unintentionally, 
purely because of the inaccurately known period.
Typical exposure time was 8 minutes corresponding to 0.015 in orbital
phases. A single spectrum at low dispersion taken in the blue
region showed that the spectral type of the star is G0V, not G5 
as given in Bond \markcite{bon76} (1976). LS~Del is a W-type W~UMa system.

LS~Del system was discovered spectroscopically as a close
binary by Bond \markcite{bon76} (1976). 
Since then, the system has been analyzed
several times. Derman \markcite{der91} (1991) 
gave a period of 0.3638384, which 
was used in our solutions of radial velocities for orbital elements.
An attempt at a solution of the light curve
(Wang et al.\ \markcite{wan86} 1986), using the mass-ratio from
the present RV data, by means of Wilson-Devinney method (Wilson 
\markcite{wil79} 1979) led to the inclination $i = 48.5$ degrees, 
almost identical temperatures of components
($T_2$ higher than $T_1$ by 40 degrees, implying a W-type system)
and a shallow contact, $f=7$\%. The absolute dimensions obtained
were: $a=2.44\,R_\odot$, $M_1=1.07\,M_\odot$ and $M_2=0.40\,M_\odot$.
Another photometric analysis of the system was carried out by Weaver
\markcite{wea90} (1990), 
who obtained a similar inclination of 
$i \simeq 45$ degree and also a weak contact; he
suggested that light variations are probably due to ellipsoidal 
deformation of components instead of eclipses; the mass-ratio was
not given. 

\subsection{EF Dra}

Two sets of spectra were obtained at DDO at two different dispersions,
10.8 \AA/mm and 30 \AA/mm. Of the total of 43 spectra, only the 16 higher
dispersion spectra (obtained in May -- June 1992) were used for the
radial-velocity determinations, which led to our first-time 
spectroscopic orbital orbit. The exposures were 20 minutes long,
corresponding to 0.033 in orbital phases.
The spectra indicated that the system is 
a triple one (Lu \markcite{lu93} 1993).
This circumstance resulted in such a heavy blending of 
the low dispersion spectra at 30 \AA/mm that no reliable
radial velocities could be measured; these spectra were not used in
our orbital determinations. The third
component is probably a physical companion, since its radial
velocity of $-38$ km/sec is very
close to the systemic velocity of $-42$ km/sec of the eclipsing pair.

The system EF Dra was discovered as an X-ray source 
by the Einstein Observatory 
Extended Medium Sensitivity Survey (Gioia et al.\ \markcite{gio87}
1987). Fleming et al.\ \markcite{fle89}
(1989) suggested it to be a W~UMa-type variable.
Robb \& Scarfe \markcite{rs89} (1989) obtained the first light curves 
and confirmed the suggestion of Fleming et al., giving the
orbital period of 0.42400 day. Later, Plewa et al.\ \markcite{ple91} (1991)
refined the period on the basis of their photometric observation.
This period was used in our orbital solution.
The photometric mass-ratio derived by Plewa et al., 
$q \simeq 0.125$ is different from our spectroscopic result
probably because they did
not take the third light into account. Our analysis shows that
EF~Dra is an A-type W~UMa system with the more massive primary component
eclipsed at the deeper minimum.

\subsection{V829~Her}

The spectroscopic orbit of V829 Her has been determined for the first
time. 30 spectra were obtained at DDO, consisting of  
26 observations during the first 
run in June 1991 and May -- June 1992 and 4 observations 
during the second run on
February 10, 1997. Of the spectra obtained during the first run,
13 (one half of the total) were secured around the orbital
conjunctions due to an inaccurately 
known period at the time of the observations. The four spectra
of the second run were all around the first orbital quadrature.
The exposure times were about 15 minutes long corresponding to
about 0.03 in orbital phases.

V829~Her was discovered serendipitously as an X-ray source during
the Einstein Observatory Extended Medium Sensitivity Survey (Gioia
et al.\ \markcite{gio87} 1987). It was suspected of being a W~UMa 
system by Fleming et al.\ \markcite{fle89} (1989).
Soon thereafter, Robb \markcite{rob89} (1989) 
obtained photometric observations
of the system and confirmed the W~UMa variability type.
He found a period of 0.35813 which was only approximate due to the short time
span. Later, he obtained a time of light minimum of 2448505.7163 (Robb
1992, private communication) which were combined
with two times of minima by 
Agerer \& Hubscher \markcite{age95} (1995) to derive
an ephemeris: Hel~Min(pri) = 2447680.8910(11) + 0.3581502(4)~E.
For this star, as for all our spectroscopic orbit solutions, 
the photometric values of periods were adopted and our
radial velocities were used only to determine the zero-epochs
of the spectroscopic orbital solutions.

By combining the observed and synthesized light curves 
(Robb R.M. 1992, private communication) and BF    
fits, in the same way as in Rucinski et al.\ \markcite{rls93} (1993),
we derived the following geometrical and
absolute parameters for this W-type W UMa system: 
$i = 57 (2)$ degree, $f = 15 (9)$\%,
$X = 0.030(6)$, $a = 2.697(13) R_\odot$, 
$R_1 = 1.266(6) R_\odot$, $R2 = 0.849(4) R_\odot$, 
$M_1=1.458(21) M_\odot$, $M_2 = 0.596(10) M_\odot$, 
and $q = 0.409(5)$.

\subsection{FG~Hya}

The spectroscopic orbit presented here has been
determined for the first time with the modern instrumentation,
although five prism spectra at a dispersion of 75 \AA/mm at
H$\gamma$ were used by Smith \markcite{smi63} (1963), from which he 
derived a primary semi-amplitude of 92 km/sec, questioned
as too large by Mauder \markcite{mau72} (1972).

The current spectroscopic orbit has been based on data obtained
at DDO in November 1996 -- February 1997.
30 spectra were secured with exposure
times of 15 minutes corresponding to 0.032 in orbital phases. 
The orbital solution assumed the period of 0.327835 following
Yang et al.\ \markcite{yan91} (1991).

After discovery of its light variation by Hoffmeister \markcite{hof34} (1934)
199), FG~Hya was the subject of many investigators.
Three sets of photometric solutions exist in the
literature: Mochnacki \& Doughty \markcite{moc72}
(1972), Twigg \markcite{twi79} (1979)
and Yang et al.\ \markcite{yan91} (1991). All these investigations found small
mass-ratios: 0.145, 0.142, 0.128, none, however, was as small as our
spectroscopic value of 0.112. We admit that for such extreme values
the spectroscopic mass-ratio may be difficult to determine accurately
and that photometric determinations utilizing eclipse contacts
may give superior results. All three photometric solutions arrived 
at a very deep over-contact; the largest 
being 90\% by Yang et al.\ \markcite{yan91} (1991). 

This A-type system deserves further investigation. Assuming the preliminary
value of $i = 87.6$ degree 
following Yang et al.\ \markcite{yan91} (1991), simply
because their mass ratio was the closest to ours from among several
photometric determinations, we derive $a=2.32 R_\odot$, $M_1=1.41 M_\odot$ 
and $M_2=0.16 M_\odot$.
These numbers are obviously approximate as the photometric solution 
was not for the same mass-ratio; for the same reason, we have not evaluated 
the radii. The point worth 
noticing is the very low mass of the secondary component. 

\subsection{AP~Leo}

The spectroscopic orbit described here has been
determined for the first time. The template-spectrum star for the CCF
determinations was HD~126053 (G1V).

A total of 21 spectra, all around at quadratures, were obtained on
three consecutive nights, February 23, 24, and 25, 1991 at 
DAO with the 1.8-m telescope and ``21121'' spectrograph with the RCA2 CCD
detector at a dispersion of 15 \AA/mm, or 0.23 \AA/pixel. All the spectra
were centered at 5020 \AA\ with a coverage of 240 \AA. 
The exposure times were 15 -- 20 minutes, corresponding to a phase
resolution about 0.024 -- 0.032. The period of 0.4303546 day
by Zhang et al.\ \markcite{zha92} (1992) was used in the spectroscopic
orbit solution. A spectrum taken in blue showed that the
spectral type of the system is F7-8V.

 Mauder \markcite{mau67}
(1967) obtained the first geometrical-elements solution, based 
on photometric data. He derived a mass-ratio of 0.31 and
inclination of 77 degree. A modern photometric
solution was carried out by Zhang et al.\ \markcite{zha92} 
(1992) who found a mass-ratio of 0.301, which is almost identical to our 
spectroscopic value. We note that a similar excellent agreement
exists for BB~Peg which was also analyzed photometrically by this group. 
However, their photometric solution of AP~Leo leads to a hotter,
less-massive secondary being eclipsed at the deeper minimum,
a configuration which is normally called the W-type 
subtype of the W~UMa systems, whereas 
our radial velocity curves show that the primary minim is a
transit, i.e.\ it is the more massive star which is 
hotter and eclipsed at the deeper minima, suggesting an A-type 
W~UMa system. This discrepancy should be clarified. There is a chance that
the photometric solution of Zhang et al.\ was affected by 
incorrect placement of spots (parenthetically, we note that the spot locations
were unbelievably accurately determined in this solution). 
Nevertheless, the inclination of 79.9 deg and fractional radii of
$r_1=0.502$ and $r_2=0.284$ (side values) derived by Zhang et al.\
could be used with some confidence to  
derive the absolute dimensions: $a=2.95\,R_\odot$, 
$M_1=1.43\,M_\odot$, $M_2=0.42\,M_\odot$, $R_1=1.48\,R_\odot$, 
$R_2=0.84\,R_\odot$. 
We note that we were not able to detect any spectroscopic
signatures of a third body in the system whose existence was suggested 
by Zhang et al.

\subsection{UV Lyn}

The spectroscopic orbit of UV Lyn has been determined by us for
the first time. The system was observed on two nights, December 29, 1990
and January 4, 1991 at DAO with the 1.8-m 
telescope and the ``21121'' spectrograph. Unfortunately, the data were
obtained with two CCD different detectors on each of the two nights.
On the first night, RCA2 (1024 pixels) was
used with a coverage of 240 \AA, while on the second night, 
FORD (512 pixels) was used with a coverage of 160 \AA.  
The dispersion was 0.23~\AA/pixel for the former detector and 
0.31~\AA/pixel for the latter. All the spectra were centered at 5020 \AA.
Two different template stars were used
during each night,  HD22484 (F8V) and HD32963 (G2V).
A total of 31 spectra were
accumulated with the exposure times ranging between 15 and 20 minutes,
with the corresponding phase resolution of 0.025 to 0.033.
RV were first measured using VCROSS and 
then re-measured using BF's. The velocities given in Table~\ref{tab1}
are those obtained by the BF method. 
The spectroscopic elements were derived by adopting the period of
0.41498088 days from  Markworth and Michaels \markcite{mar82} (1982).
 
The variability of UV~Lyn was discovered by Kippenhahn 
(Geyer et al.\ \markcite{gey55} 1955). There are three sets of photoelectric 
observations in the literature, with the most recent one 
by Zhang et al.\ \markcite{zha95} 
(1995) who published complete light curves and listed
two times of minima. Earlier photometric
data sets were obtained by Bossen \markcite{bos73} (1973) and
by Markworth and Michaels \markcite{mar82} (1982); both groups attempted
photometric solutions, but derived photometric mass-ratios 
very different from those presented here so that 
other parameters of these solutions are obviously highly questionable. 

An attempt was made to produce a combined solution of geometrical
elements of UV~Lyn, based on
the light curve by Bossen \markcite{bos73} (1973) and the BF's obtained
here. The light curve by Bossen was used\footnote{The light curve 
was made available by the librarian P.D.\ Hingley at the
Royal Astronomical Society, London, UK.}, since 
the Markworth and Michaels light curves were very perturbed.

The approach consisted of fitting the theoretical BF's computed
with the light-curve/line-profile synthesis program
WUMA5 (a very brief infomation in Hill \& Rucinski \markcite{hr93} 1993
and in a short series of papers cited below)
to the observed BF's, using the spectroscopic
mass ratio $q$ derived here as the initial value.
A similar approach was used in the three papers
of a short series: Rucinski \markcite{ruc92} 1992,
Lu \& Rucinski \markcite{lr93} 1993,
Rucinski et al.\ \markcite{rls93} 1993).
The mass-ratio changed from the spectroscopic value
insignificantly to $q = 0.371(8)$, with the remaining parameters:
$i = 67 (2)$ deg, $f = 20$\%, $X = 0.018 (4)$.
The absolute parameters are: $a = 2.914(17)$ $R_\odot$,
$M_1=1.412(31) M_\odot$, $M_2=0.523(15) M_\odot$, 
$R_1 = 1.403(8) R_\odot$, $R_2 = 0.905(5) R_\odot$.

\subsection{BB Peg}
 
The present spectroscopic-orbit determination is the second one,
after the solution by Hrivnak \markcite{hri90} (1990). Our observations were
obtained at DDO in August -- September 1996.
28 spectra were obtained with exposure times of 15 minutes 
corresponding to 0.029 in orbital phases.
The orbital period in our solution was fixed at the value of 0.361501 
determined by Leung et al.\ \markcite{leu85} (1985).

The system was discovered by Hoffmeister \markcite{hof31} (1931)
and since then has been the topic of several investigations. 
The only spectroscopic orbit solution of 
Hrivnak \markcite{hri90} (1990) was based on radial velocity data
obtained from spectral cross-correlations.
His mass-ratio, $q = 0.34 \pm 0.02$, is in a good agreement with ours,
which was derived from the BF's. The mass-ratio derived from CCF
method is not too significantly different from the one from BF method, but the
cross-correlation method carries a potential of underestimating
the amplitude of the more massive star, which would lead 
to a smaller value of $q$. The most consistent approach in the derivation of
physical parameters is a combined solution of photometric and spectroscopic
data, as in the case of AH Vir (Lu and Rucinski \markcite{lr93}
1993) and also in the case of UV Lyn (this paper),
which in both cases confirmed the BF result. 
As we understand Dr.\ Hrivnak plans to obtain a combined 
photometric and spectroscopic solution on the basis of his CCF
determination of $q$. This will provide an external check on our
solutions as the two datasets are both of good quality but
also entirely independent of each other.

Several photometric solutions of light curves are available, among them
by Cerruit-sola et al.\ \markcite{cer81} (1981). They
derived a mass-ratio 2.49 or, following our convention, $q = 0.40$, 
which is significantly different from ours. Moreover,
they found a rather deep contact configuration of 37\%, which is
uncommon for W-type W~UMa stars. Leung et al.\ \markcite{leu85} (1985) 
solved their light curves of BB~Peg and derived a mass-ratio of 
0.356, which is almost identical to our spectroscopic $q$.  
They also found a more likely fill-out factor of 12\%. 
Using their geometrical parameters (inclination and fractional
radii), the following absolute dimension of this W-type system have been 
derived: $a=2.63 \, R_\odot$, side radii $R_1=1.26 \, R_\odot$
and $R2=0.76 \,R_\odot$, $M_1=1.38 \, M_\odot$, $M_2=0.50 \, M_\odot$.

\subsection{AQ Psc}

The spectroscopic orbit of this system is presented for
the first time. It was observed at DDO, in August -- October 1996.
Due to the inaccurately known
period, from among our 37 spectra, 11 were secured around
conjunctions rather than quadratures. The exposure times were
15 minutes long, corresponding to about 0.022 in orbital phases.

This system has not been investigated extensively before. Pastori 
\markcite{pas85} (1985)
discovered it to be a W~UMa variable and gave a period of 0.47564 day. Since
then a few more times of minima were available (Muyesserroglu et al.\
\markcite{muy96} 1996, Demircan, O., 1997, private communication).
With these limited material available, we derived a new ephemeris:
Hel~Min (pri) = 2449283.3292(16) + 0.4756056(4), of which only the
period was used in the spectroscopic orbit solution. Although the period was
not precisely determined, it should be accurate enough for our spectroscopic
solution since the time span of our observations was relatively short. 
With our well-determined spectroscopic orbit, this A-type system deserves a
new photometric investigation. 

\section{Summary}

The paper presented
the first installment of radial velocity determinations for ten bright,
close-binary systems lacking adequate radial velocity data.
The new results have been obtained using the modern techniques of
template-matching (cross-correlation and broadening-function) 
on the basis of observations collected with the
1.8-meter class telescopes at the Dominion Astrophysical Observatory
and David Dunlap Observatory. Preliminary, circular-orbit
spectroscopic orbit solutions have been obtained for all ten systems
with the immediate goal of extracting the ``gamma'' velocities
and mass-ratios. The former will be used for a re-discussion of
the kinematics of the contact systems while the latter
can serve as an important departure point in combined, precise 
photometric/spectroscopic synthesis solutions of individual systems.

\acknowledgements

The authors would like to thank the directors of Dominion 
Astrophysical Observatory and of David Dunlap Observatory for 
generous allocation of observing time.

\clearpage

\noindent
Captions to figures:

\bigskip

\figcaption[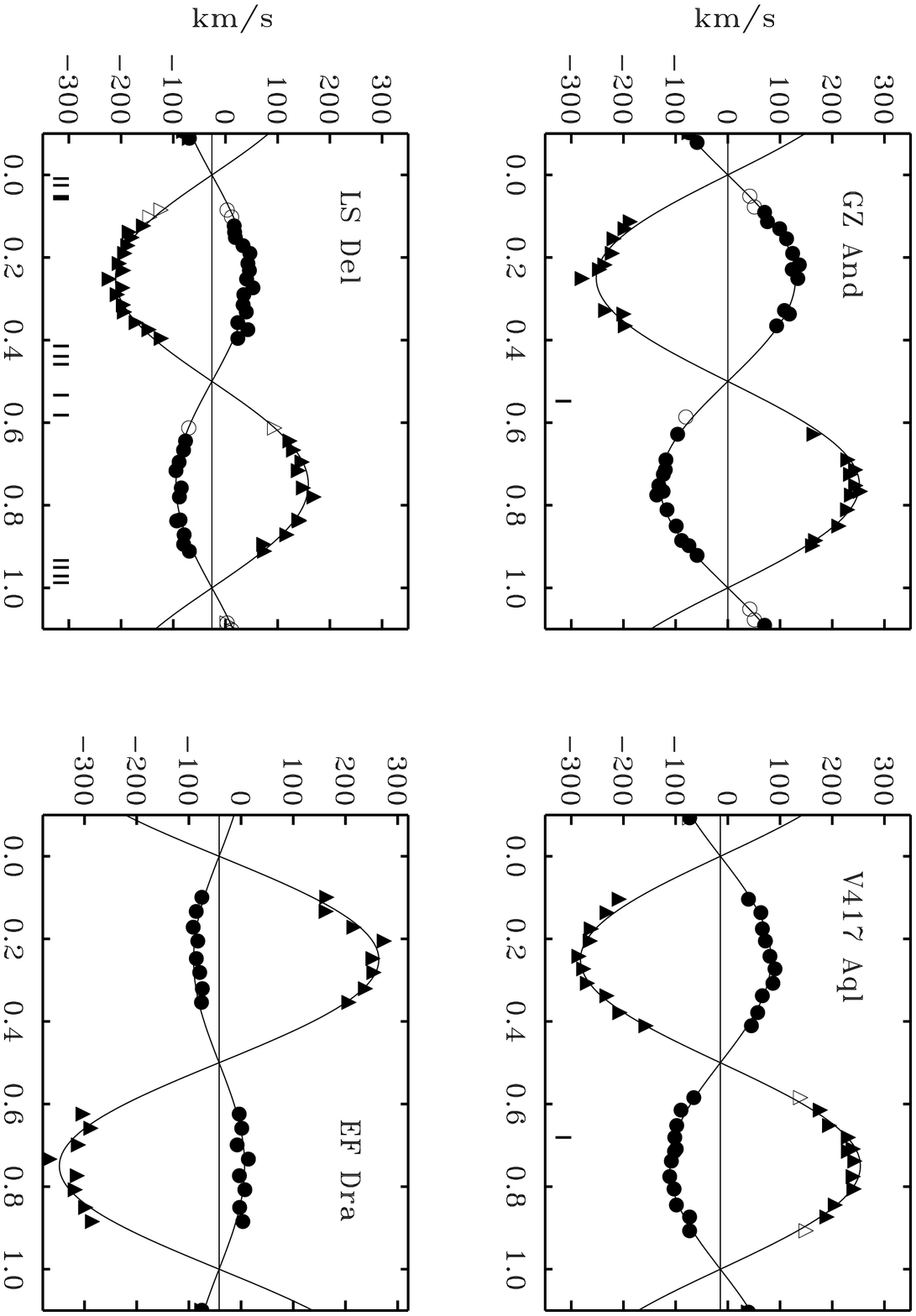] {\label{fig1}
Radial velocities of the first four
systems, GZ~And, V417~Aql, LS~Del, EF~Dra, are plotted in individual
panels versus orbital phases. The thin lines give the
respective circular-orbit (sine-curve) fits to the radial velocities.
Open symbols indicate observations given half-weights in the solutions
while marks at the lower show phases of available observations which
were not used in the solutions because of the 
blending of lines. EF~Dra is the only 
A-type system among the four systems shown here;
all other systems are of the W-type with slightly hotter less-massive
components.
}

\figcaption[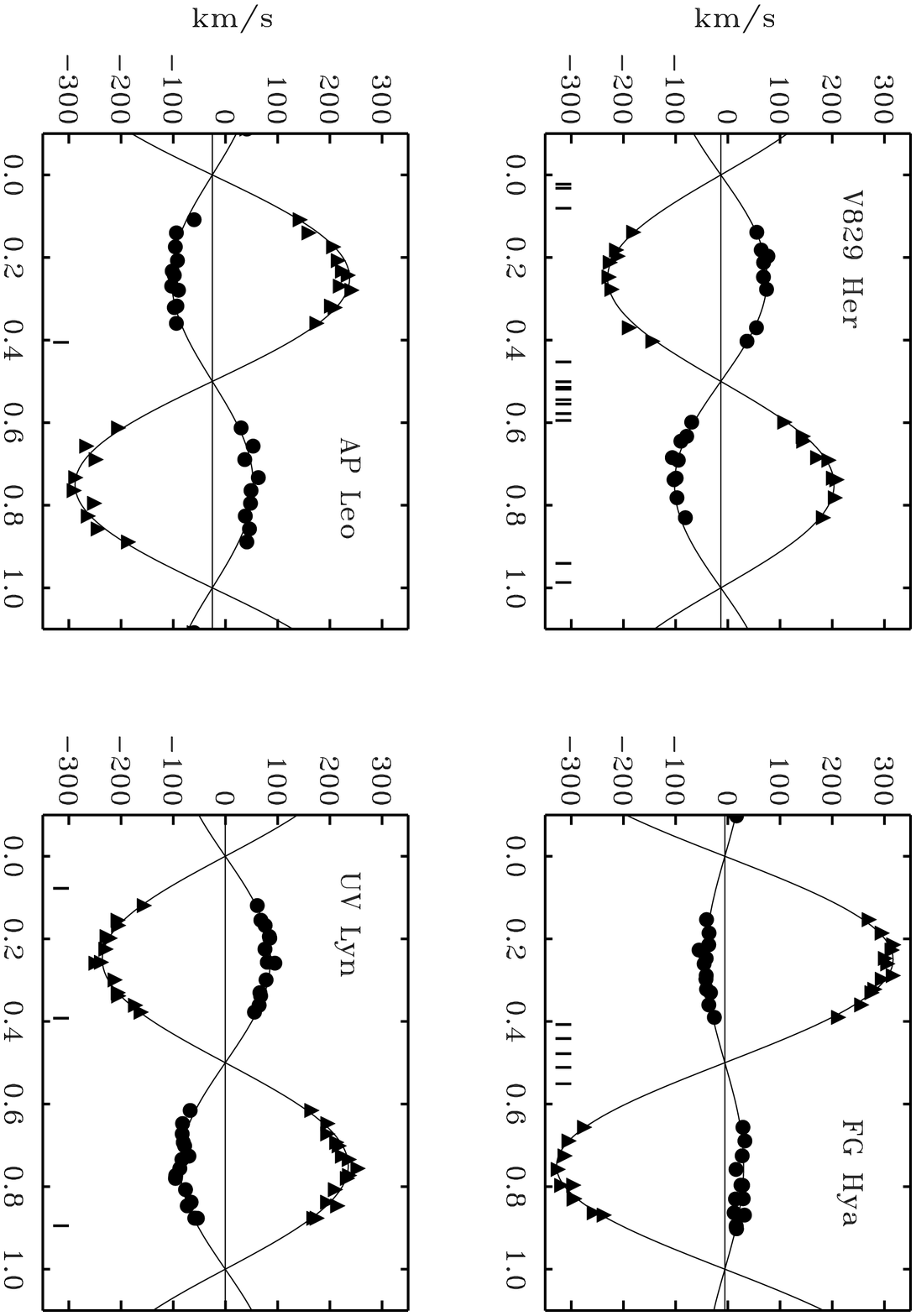] {\label{fig2}
Radial velocities of the systems V829~Her, FG~Hya, AP~Leo and UV~Lyn,
in the same format as in Figure~1. Among the four systems,
FG~Hya and AP~Leo belong A-type.
}

\figcaption[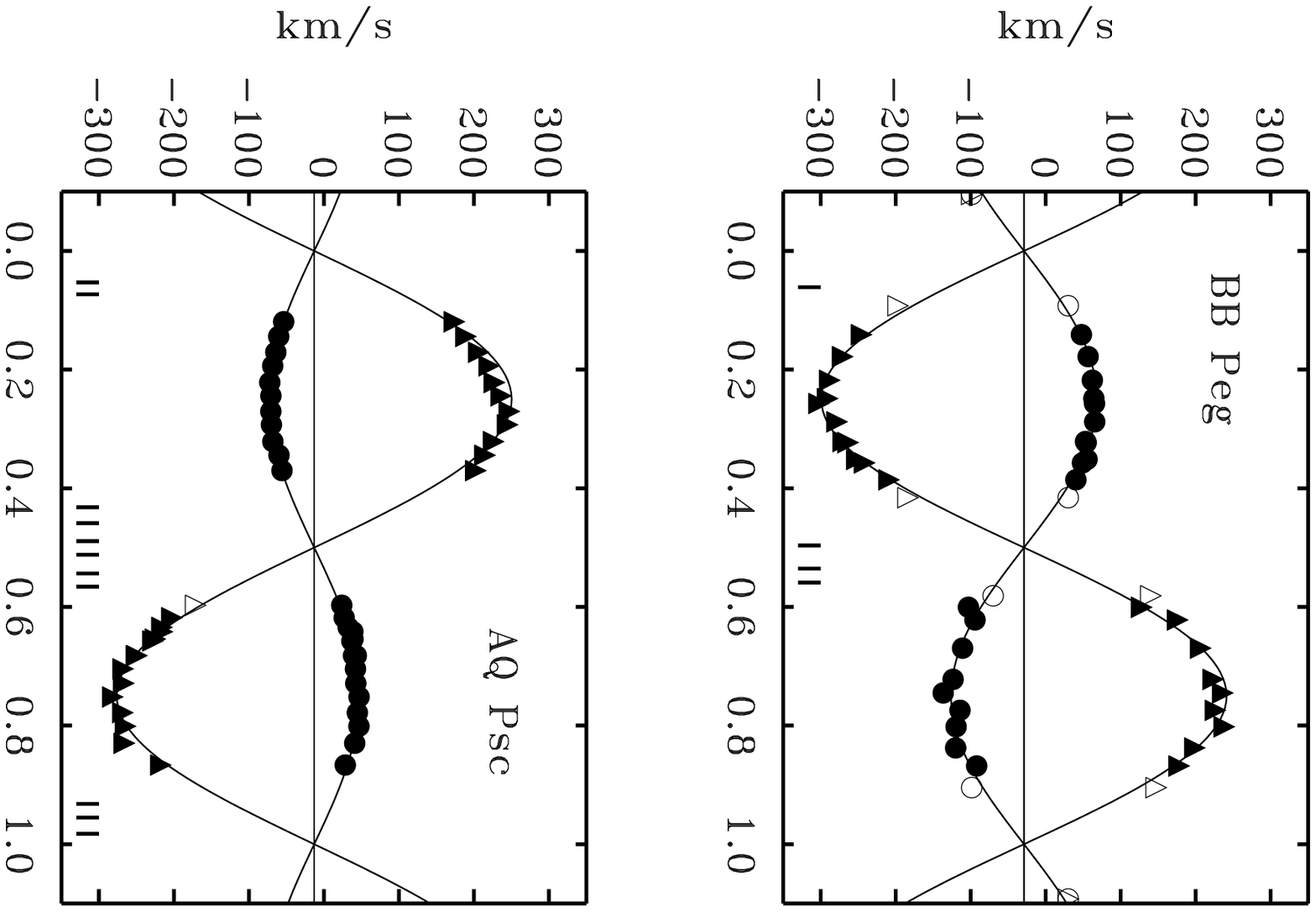] {\label{fig3}
Radial velocities of the systems BB~peg and AQ~Psc,
in the same format as in Figures~1 and 2.  AQ~Psc is an A-type system.
}

\clearpage

\begin{table}                 
\dummytable \label{tab1}      
\end{table}

\begin{table}                 
\dummytable \label{tab2}      
\end{table}

\end{document}